\documentstyle[12pt,leng]{article}
\begin{document}
{\flushright IP-ASTP-16-93}\\
{\flushright October 7, 1993}
\vspace{24pt}
\begin{center}
{\large\sc{\bf Edge states in strong magnetic field.}}
\baselineskip=12pt
\vspace{35pt}

I. Barto\v{s}\footnote{permanent address: Institute of Physics,
Academy of Sciences of the Czech Republic, Prague} and B.Rosenstein
\vspace{24pt}

 Institute of Physics \\
Academia Sinica\\
Taipei, 11529\\
Taiwan\\
\vspace{60pt}
\end{center}
\baselineskip=24pt
\begin{center}
{\bf abstract}
\end{center}

Energies and wave functions of edge states in two
dimensional electron gas
are evaluated  for a finite step
potential barrier model. The spectrum exhibits richer structure than observed
previously
for an infinite barrier model. Surprisingly,  instead of
smooth Landau level bending in the vicinity of the barrier,
the levels acquire a steplike form. These plateaus have a direct impact on
 widths  of  the magnetotransport conducting channels.

\vspace{10pt}
\vspace{30pt}

\pagebreak

Strong magnetic fields allowed the discovery of
such dramatic phenomena in two dimensional electron gas (2DEG)
like Integer
and Fractional Quantum
Hall Effect. In the microscopic theory \cite{PG} the edge states,
classically represented by electrons skipping in circular segments
along the edges, play
a dominant role. Many
magnetotransport experiments in two dimensional electron gas have been
qualitatively understood recently by means of a simple edge - state
model \cite{H}.
This model is based on the picture of smooth Landau level bending
by the potential formed by external charges. The intersections
of each Landau level with the Fermi surface create widely
separated narrow edge channels \cite{Ha}. Recently Chklovskii {\it et al}
\cite{C}, using the self-consistent electrostatic approach, showed that
the resulting effective potential should acquire a steplike shape.
The Landau levels
 bending following these steps would transform the narrow edge channels
into broader ones.

The semiclassical notion of the Landau levels bending is not always
applicable to the electronic structure
in the vicinity of the barrier representing an edge or
an interface (or
random potential). Existing explicit
quantum mechanical model \cite{MDS} is limited to the extreme case of an
infinite barrier. This special barrier may fail to
account for certain aspects of the surface/interface electronic
structure, e.g. the existence of localized
surface/interface states \cite{B}.

In this paper, finite step potential is considered to better approximate the
interface. This might be either be a "boundary" confining 2DEG or an
interface between two different materials. The finiteness of the
 barrier confining 2DEG is especially important if the confinement is
realized by means of an interface between two similar materials.
Then, in very strong magnetic fields the induced level splitting may reach
comparable values.

Here, we study a system of 2D noninteracting electrons in the vicinity of a
boundary under the homogeneous
magnetic field  $B$ perpendicular to the  $xy$ - plane.
The edge or interface is described by a finite potential barrier between two
materials (regions) L and R.   In the
region L the potential is lower by an amount $V$
 than in the region R.
 We assume that the interface is placed
along the line
$x=0$.

The one particle Hamiltonian in the Landau gauge ${\vec A}\equiv (0,Bx,0)$
 is
\begin{equation}
H=\frac{1}{2m} ({\vec p} - \frac{e}{c} {\vec A})^2+V\theta(x).
\end{equation}
In this gauge
the motion along the y - direction is free and
we can separate variables:
\begin{equation}
\psi_{n,X} (x,y)=\frac{1}{{\sqrt 2\pi}}{\exp \left(\frac{iXy}{a_L^2}\right)}
\phi_{n,X}(x)
\end{equation}
where $\phi_{n,X}(x)$ obeys the one dimensional equation
\begin{equation}
\left[-\frac{\hbar}{2m}\frac{d^2}{dx^2}+\frac{1}{2}m\omega_c^2(x-X)^2+
V\theta(x)\right]
\phi_{n,X}(x)=E_{n,X}\phi_{n,X}(x)
\end{equation}
where $a_L^2\equiv \frac{\hbar c}{eB}$ is the magnetic length, $\omega_c\equiv
\frac {eB}{mc}$ is the cyclotron frequency and $X$ is the distance of the
center of a Larmor orbit from the interface.
This is  the Schr\"odinger equation
for the harmonic oscillator superimposed with the step barrier.
The integer
$n$ parametrizes discrete Landau levels.
In the extreme
case of an infinite step barrier,  simple boundary condition
of vanishing of the wave function is imposed:
\begin{equation}
\phi_{n,X}(0)=0
\end{equation}
 It was studied by McDonald and Streda \cite{MDS},
who obtained energies of first few levels as a function of
the distance $X$. Deep inside the region L ($X<<0$) the influence
of the interface is negligible and $E_{n,X}\rightarrow \hbar \omega_c \left(n
+\frac {1}{2}\right)$, $n$ counting the
Landau levels in the bulk. As $X$ approaches the barrier,
 the energy levels rise
due to repulsive effect of the infinite barrier. For orbits centered
in the "forbidden" region R the energies continue to rise indefinitely.

In systems of practical importance, the potential barrier is typically {\it
not}
much larger than the magnetic level spacing $\hbar \omega_c$.
For example, in the 2DEG formed at the interface between GaAs and GaAl$_{1-x}$
As$_x$, the effective interface potential barrier is only about $0.3 eV$.
The Landau level spacing in strongest magnetic fields experimentally available
is just a few times smaller.

When the interface barrier is finite, simple boundary condition
eq.(4) should be replaced by a quantum mechanical matching of general
solutions of eq.(3) along the boundary line $x=0$. Because of the explicit
translational
invariance in the $y$ direction, the wave function matching for $\phi_{n,X}(x)$
has to be performed at a single point $x=0$.
It is convenient to shift the origin of the coordinate system to $X$; in
 natural
units of magnetic length: $x'\equiv  \frac{\sqrt 2}{a_L} (x-X)$. Energy
expressed
in units of Landau spacing $\hbar \omega_c$ is
$\epsilon_n\equiv E_{n,X}/(\hbar \omega_c) \equiv \nu_n+\frac {1}{2}$. Equation
(3) then takes a form
\begin{equation}
\left[ \frac {d^2}{dx'^2}-\frac {1}{4} x'^2 - \frac{V}{\hbar \omega c}
\theta(x'+X) +
\left(\nu_n+\frac {1}{2}\right)
\right]\phi'_{n,X}(x')=0
\end{equation}
which is the differential equation defining the parabolic cylinder functions
\cite{E}. The two linearly independent solutions $D_{\nu_n}(x')$ and
$D_{\nu_n}(-x')$ satisfy asymptotically the conditions of  rapid decrease
for $x'\rightarrow +\infty$ and  $x'\rightarrow -\infty$ respectively.

The matching of the logarithmic derivatives at $x'=-X$ (or equivalently
the condition of zero Wronskian in the expression for the Green's
function of the system) gives
\begin{equation}
\frac {D'_{\nu_n-V}(-x')|_{x'=-X}}{D_{\nu_n-V}(-X)}+
\frac {D'_{\nu_n}(x')|_{x'=-X}}{D_{\nu_n}(-X)}=0
\end{equation}
This determines the energy levels $\nu_n$ as functions of the position $X$.
The equation (6) was solved numerically using a simplified form \cite{pc}
\begin{equation}
 D_{\nu_n+1}(-X) D_{\nu_n-V}(X)+ D_{\nu_n-V+1}(X) D_{\nu_n}(-X)=0
\end{equation}

Eigenvalues $\nu_n$
for a few lowest Landau levels have been evaluated as functions of
$X$, the center of a Larmor orbit if the interface is not interfering,
for two values of the interface potential $V$ and are compared
with those for an infinite wall \cite{MDS}, see Fig.1.
Only the lowest three levels are shown. For small barrier $V= 1/2 \ \ \hbar
\omega_c$ (three bottom curves) the levels simply bend a little.
For an infinite barrier (three top curves)
the levels continue to rise to infinity.
The energies of states localized deep in the region L ($X<<0$) are not
affected by the interface. They form the bulk Landau bands at $\nu_n=n$
and asymptotically do not depend on $X$. Qualitatively, when the
radius of the classical Larmor orbit approaches the interface its
repulsive effect pushes the energy upwards. Consequently at
a given localization
$X$, higher Landau levels start deviating from their bulk
energies earlier then the lower ones due to their larger size $\simeq
{\sqrt n}a_L$.

Energies gradually saturate as the "orbit center" $X$ moves deeper into the
region R. Asymptotically, for $X>>0$ they reach values shifted by amount $V$
with respect to those in the bulk of the region A.
The transition over the interface is accomplished slowlier for higher Landau
levels due again to their larger size.

The staircase shape of the curves for $V=5 \hbar\omega_c$ (three middle curves
on Fig.1) can be understood as
a tendency of each Landau level from region L to occupy the
lowest level in R . The repulsion from the lower states pushes it towards the
next
plateau due to the noncrossing rule for electron levels. Note that the
neighbouring plateaus are separated by very small gaps. We found
numerically that the minimal gap between lowest Landau level and first excited
one is just 0.12 $\hbar \omega_c$.

Because of the nature of the mechanism just described \cite{f}, the staircase
type of transition from L to R is not restricted to
 just the finite step potential.  In fact, similar structures had been
observed some time ago for a different potential. Heinonen and Taylor studied
Hartree type self consistent solutions \cite{HT}. Their Fig. 2 indicates
somewhat blurred features of plateaus.

Flat regions do not contribute to electric
current and therefore spatially separated conducting edge channels are
formed, similarly as in \cite{C} although the mechanism is different from this
in \cite{C} and
similar to that of Heinonen and Taylor \cite{HT}.

In contrast to the infinite
potential barrier when electrons are restricted to the region L only,
here the wave functions penetrate into the region R. Far from the interface
in the region R when tunneling is negligible  they restore their
 bulk shapes.
Several typical probability densities of the states in the lowest and
the first excited Landau levels in the vicinity of the
step barrier are shown in Fig.2 and 3 respectively.

The ground state wave function, centered far from the barrier ($X=-2.12$,
the leftmost curve in Fig.2) is a Gaussian. As $X$ approaches the barrier, the
wave function gets compressed. Two highest peaks, corresponding to orbits
"centered" on the barrier ($X=0.71$ and $2.12$) clearly demonstrate how the
centers of mass are lagging behind their classical
Larmor orbit centers $X$. Only when the energy of a state reaches
its saturation value inside B, the Gaussian shape is recovered ($X= 3.54$,
the rightmost curve in Fig.2).

The next Landau level transformations in the region of the first
energy plateau \newline
($\nu\simeq 5 \hbar \omega_c$, $X$ around 2)
are illustrated in Fig. 3. Two dashed curves ($X=1.77 $ and $X=2.12$)
are nearly Gaussians with only a small bump to the left from their
node. When the orbit "center" is moved deeper into R ($X=2.48$, the
solid line), after its energy is raised correspondingly, the electron
center of mass slips back into the region L left of the barrier.
In this case, not only the center of mass lags behind $X$, but even moves
in the opposite direction.

The fact that the center of the electron's wave function lags behind
the orbital center $X$ when the latter is moved deeper into the region R is
sometimes
used  as an argument for enhanced electric charge density at the edge\cite{H}.
However this lagging behind is not the only physical effect determining
the charge density balance. Another relevant effect is the increased
energy of the states which can become depopulated while crossing the
Fermi level. Obviously, a proper self  - consistent quantum mechanical
calculation is
needed to quantitatively determine the overall balance.

To summarize, energies of electron states approaching in the strong
perpendicular
magnetic field the finite step potential barrier, representing edge
or interface
 in the 2DEG, have been investigated. In contrast to the infinite barrier
model,
the energies rise in steps (Fig.1). The $n^{th}$ Landau level
exhibits $n-1$ flat plateaus corresponding to  $n-1$
lowest Landau levels in medium R. There is a curious "apparent level crossing"
in a sense that higher level approaches very close the lower ones.
Flat regions do not contribute to electric
current and therefore spatially separated conducting edge channels are
formed, similarly as in \cite{C} though the mechanism of their formation is
completely different. Here we obtain the plateaus strictly as a one electron
effect without any corresponding structure of the potential.

We acknowledge the support of  National Science Council ROC, grants
NSC-82-0208-M-001-116 (B.R.) and NSC-82-0501-I-00101-B11 (I.B.).

\newpage

FIGURE CAPTIONS\newline
\newline

Fig. 1

Eigenvalues of three lowest Landau levels, $E_{n(X)}
=\hbar \omega_c(n(X)+1/2)$. The energies are  expressed in terms of $n(X)$,
as functions of $X$, the Larmor's orbit center with respect to the barrier.
Different curves correspond to three heights of the potential barrier:
$V=0.5  \ \ \hbar \omega_c$ (three  lines at the bottom), $V=5  \ \ \hbar
\omega_c$ (three lines in the middle) and
$V=\infty$ (three lines at the top).
\newline\newline

Fig. 2

Electron density $|\phi_0(x)|^2$ for the lowest
 Landau level in the vicinity of the
potential barrier. Larmor orbits are centered at five successive equidistant
 locations: $X=$-2.12 (small dashes), $X=$-0.71 (large dashes), $X=$0.71
(dot-dash),  $X=$2.12
(small dashes), $X=$ 3.54 (solid line).
\newline\newline

Fig. 3

Electron density $|\phi_1(x)|^2$ for the first excited
 Landau level in the vicinity of the plateau of Fig.1.
Larmor orbits are centered at three successive equidistant
 locations: $X=$ 1.77 (dashes),  $X=$ 2.12 (dot-dashed),  $X=$ 2.48 (solid
line).


\newpage


\begin{thebibliography}{99}
\bibitem{PG}  R.E. Prange and S.M. Girvin, {\it The Quantum Hall Effect},
New York, Springer, 1987.
\bibitem{H} R.J. Haug, {\it Semicond. Sci. Technol.} {\bf 8}, 131 (1993)
and references therein.
\bibitem{Ha} B. I. Halperin,  {\it Phys. Rev} {\bf B25}, 2185 (1982).
\bibitem{C} D.B. Chklovskii, B.I. Shklovskii and L.I. Glazman,
 {\it Phys. Rev} {\bf B46}, 4026 (1992).
\bibitem{MDS} A.H. MacDonald and P. St\v{r}eda, {\it Phys. Rev} {\bf B29},
 1616 (1984).
\bibitem{B} B. Velick\'y and I. Barto\v{s}, {\it J. Phys. C: Solid State Phys.}
{\bf 4}, 19 (1971).
\bibitem{E} A. Erdelyi. et al. , {\it Higher Transcendental Functions},
McGraw Hill, New York, 1954; E. Merzbacher, {\it Quantum Mechanics}, Chapter 5,
John Wiley, New York, 1970.
\bibitem{pc}Here
well known relations expressing derivatives of the parabolic cylinder
functions \cite{E} were used.
\bibitem{f} The integer value of the barrier is not crucial to
obtain the plateau structure: we observed similar structure
for other values of $V$, too.
\bibitem{HT} O. Heinonen and P.L. Taylor,  {\it Phys. Rev} {\bf B32},
633 (1985).

\end{thebibliography}
\end{document}